\documentclass[longauth,traditabstract]{aa} 
\usepackage{graphicx}
\usepackage{txfonts}
\begin{document}
   \title{Detection of hydrogen fluoride absorption in diffuse molecular clouds 
with {\it Herschel}/HIFI: a ubiquitous tracer of molecular gas.\thanks{{\it Herschel} is an ESA space observatory with science instruments provided
by European-led Principal Investigator consortia and with important participation from NASA.}}


   \author{P.~Sonnentrucker\inst{1}, D. A. Neufeld\inst{1}, T.~G.~Phillips \inst{2}, M.~Gerin \inst{3}, D.~C.~Lis \inst{2},  M.~De~Luca \inst{3}, J.~R.~Goicoechea \inst{4}, J.~H.~Black \inst{5}, T.A.~Bell\inst{2},  F.~Boulanger \inst{6}, J.~Cernicharo \inst{4}, A.~Coutens\inst{7}, E.~Dartois \inst{6}, M.~Ka{\'z}mierczak\inst{8}, P.
  ~Encrenaz \inst{3}, E.~Falgarone \inst{3}, T.~R.~Geballe \inst{9}, T.~Giesen\inst{10}, B.~Godard\inst{3},  P.~F.~Goldsmith \inst{11}, C.~Gry \inst{12}, H.~Gupta\inst{11},  P.~Hennebelle \inst{3}, E.~Herbst \inst{13}, P.~Hily-Blant\inst{14}, C.~Joblin\inst{7},  R.~Ko{\l}os \inst{15}, J.~Kre{\l}owski\inst{8}, J.~Mart\'in-Pintado\inst{4},  K.~M.~Menten\inst{16}, R.~Monje\inst{2}, B.~Mookerjea \inst{17}, J.~Pearson \inst{11}, M.~Perault \inst{3}, C.~M.~Persson \inst{5}, R.~Plume \inst{18}, M.~Salez \inst{3}, S.~Schlemmer\inst{10}, M.~Schmidt\inst{19}, J.~Stutzki \inst{10}, D.~Teyssier\inst{20}, C.~Vastel \inst{7},  S.~Yu \inst{11}, E.~Caux\inst{7}, R.~G{\"u}sten\inst{16}, W.~A.~Hatch\inst{11}, T.~Klein\inst{16},  I.~Mehdi\inst{11}, P.~ Morris\inst{21} and J.~S.~Ward\inst{11}}

   \institute{The Johns Hopkins University, Baltimore, MD 21218, USA \\
 \email{sonnentr@pha.jhu.edu}
\and California Institute of Technology, Pasadena, CA 91125, USA
\and LERMA, CNRS, Observatoire de Paris and ENS, France
\and Centro de Astrobiolog\`{\i}a, CSIC-INTA, 28850, Madrid, Spain
\and Chalmers University of Technology, G\"oteborg, Sweden
\and Institut d'Astrophysique Spatiale (IAS), Orsay, France
\and Universit\'e Toulouse; UPS ; CESR ; and CNRS ; UMR5187,
9 avenue du colonel Roche, F-31028 Toulouse cedex 4, France 
\and Nicolaus Copernicus University, Tor{\'u}n, Poland
\and Gemini telescope, Hilo, Hawaii, USA
\and I. Physikalisches Institut, University of Cologne, Germany
\and JPL, California Institute of Technology, Pasadena, USA 
\and LAM, OAMP, Universit\'e Aix-Marseille \& CNRS, Marseille, France
\and Depts.\ of Physics, Astronomy \& Chemistry, Ohio State Univ.\, USA
\and Laboratoire d'Astrophysique de Grenoble, France
\and Institute of Physical Chemistry, PAS, Warsaw, Poland
\and MPI f\"ur Radioastronomie, Bonn, Germany
\and Tata Institute of Fundamental Research, Homi Bhabha Road, Mumbai 400005, India
\and Dept. of Physics \& Astronomy, University of Calgary, Canada
\and Nicolaus Copernicus Astronomical Center, Tor{\'u}n, Poland
\and European Space Astronomy Centre, ESA, Madrid, Spain
\and Infrared Processing and Analysis Center, California Institute of Technology, MS 100-22, Pasadena, CA 91125}
  
   \date{Preprint online version: May 25, 2010}

 
  \abstract
{We discuss the detection of absorption by interstellar hydrogen fluoride (HF) along the sight line to the submillimeter continuum sources W49N and W51.  We have used {\it Herschel}'s HIFI instrument in dual beam switch mode to observe the 1232.4762 GHz $J = 1 - 0$ HF transition in the
upper sideband of the band 5a receiver.  We detected foreground absorption by HF toward both sources over a wide range of velocities.  Optically thin absorption components were detected on both sight lines, allowing us to {\it measure} -- as opposed to obtain a lower limit on -- the column density of HF for the first time.  As in previous observations of HF toward the source G10.6--0.4, the derived HF column density is typically comparable to that of water vapor, even though the elemental abundance of oxygen is greater than that of fluorine by four orders of magnitude.   We used the rather uncertain $N$(CH)-$N$(H$_2$) relationship derived previously toward diffuse molecular clouds to infer the molecular hydrogen column density in the clouds exhibiting HF absorption.   Within the uncertainties, we find that the abundance of HF with respect to H$_2$ is consistent with the theoretical prediction that HF is the main reservoir of gas-phase fluorine for these clouds. Thus, hydrogen fluoride has the potential to become an excellent tracer of molecular hydrogen, and provides a sensitive probe of clouds of small H$_2$ column density.  Indeed, the observations of hydrogen fluoride reported here reveal the presence of a low column density diffuse molecular cloud along the W51 sight line, at an LSR velocity of $\sim 24\, \rm km\,s^{-1}$, that had not been identified in molecular absorption line studies prior to the launch of {\it Herschel}.}

\authorrunning{Sonnentrucker et al.}
\titlerunning{HF in diffuse molecular clouds}
   \maketitle
%

\section{Introduction}
In some of the first results obtained using the HIFI instrument (De Graauw et al.\ 2010) on {\it Herschel} (Pilbratt et al.\ 2010), the detection of hydrogen fluoride has been reported along sight lines to the bright continuum source G10.6--0.4 (Neufeld et al.\ 2010) and the Orion hot core (Phillips et al.\ 2010).  In addition, absorption by HF has been detected (although spectrally unresolved) in the spectrum of G29.96--0.02, even at the lower spectral resolution of the SPIRE instrument (Kirk et al.\ 2010).  For G10.6--0.4, observations of the $J=1-0$ transition of HF revealed optically thick absorption in foreground gas clouds unassociated with the continuum source, and implied a lower limit of 30$\%$ for the fraction of gas-phase fluorine nuclei in HF.  Remarkably, the inferred hydrogen fluoride abundance was comparable to that of water vapor, even though the interstellar fluorine abundance is four orders of magnitude lower than that of oxygen.  These observations corroborated a theoretical prediction (Neufeld, Wolfire \& Schilke 2005; Neufeld \& Wolfire 2009) that HF is the dominant reservoir of fluorine wherever the interstellar H$_2$/atomic H ratio exceeds $\sim 1$; the unusual behavior of fluorine is explained by its unique thermochemistry, F being the only atom in the periodic table that can react exothermically with H$_2$ to form a hydride.  An implication of this theoretical prediction, together with the detection of strong HF absorption along the sight lines to G10.6--0.4, the Orion hot core, and G29.96--0.02, is that HF may serve as a valuable surrogate tracer for molecular hydrogen within the diffuse interstellar medium, both in the Milky Way and other galaxies.   

As part of the PRISMAS (PRobing InterStellar Molecules with Absorption line Studies) Key Program, we have observed hydrogen fluoride toward two additional Galactic continuum sources with sight lines that intersect foreground gas clouds: the star-forming regions W49N and W51, located at distances of $\sim 11.4$ (Gwinn et al.\ 1992) and $\sim 5.4$~kpc (Sato et al.\ 2010) from the Sun, respectively.  Previous observations of both sources have led to the detection of foreground absorption by many species, including atomic hydrogen (e.g. Fish et al.\ 2003), OH, H$_2$O (Neufeld et al.\ 2002; Plume et al.\ 2004), HCO$^+$, CN, HCN, and HNC (Godard et al.\ 2010).  In general,  many of the molecular absorption features observed previously toward W49N, and particularly toward W51, are less saturated than those measured toward G10.6--0.4; this difference makes W49N and W51 attractive targets for observations of HF absorption, because our previous study of G10.6--0.4 yielded only a lower limit on the HF abundance, as the observed absorption line was completely saturated over most velocities.  In this letter we present the results of observations of HF $J=1-0$ obtained toward W49N and W51.


\section{Observations and data analysis}


 The $J=1-0$ transition of HF, with rest frequency 1232.4762 GHz (Nolt et al.\ 1987), was observed toward W49N and W51 on 2010 March 22 in the upper sideband of HIFI receiver band 5a.  The observations were carried out at three different local oscillator (LO) tunings in order to securely identify the HF line toward both sight lines. The dual beam switch mode (DBS) was used with a reference position located 3$^{\prime}$ on either side of the source position along an East-West axis. We centered the telescope beam at $\alpha =$19h10m13.2s, $\delta =$ 09$^{\circ}$06$^{\prime}$12.0$^{\prime\prime}$ for W49N and $\alpha =$19h23m43.9s, $\delta =$ 14$^{\circ}$30$^{\prime}$30.5$^{\prime\prime}$ for W51 (J2000.0). The total on-source integration time amounts to 222s on each source using the Wide Band Spectrometer (WBS) that offers a spectral resolution of 1.1 MHz ($\sim$0.3 km s$^{-1}$ at 1232 GHz). 

The data were processed to Level 2 using the standard HIFI pipeline and HIPE version 2.4 (Ott 2010), hence providing fully calibrated spectra for both polarization modes at each LO setting. We further analyzed the Level 2 data using IDL routines developed locally. We found that the signals measured in the two polarization modes for each LO setting were in excellent agreement, as were the spectra obtained at  the three LO tunings. We therefore produced an average spectrum that consists of the weighted sum of all six observations (three LO tunings with two polarization modes each), where each observation is weighted in proportion to its signal-to-noise ratio.  The resulting double sideband continuum antenna temperatures are $T_A$(cont)$=$ 10.65 K for W49N and $T_A$(cont)$=$ 10.17 K for W51.  The r.m.s. noise was 0.13 K for both sources.

HIFI employs double sideband receivers. Thus for a sideband gain ratio of unity, the complete absorption of radiation at a single frequency will reduce the measured antenna temperature to one-half the apparent continuum level.  Assuming the sideband gain ratio is equal to unity,  the flux normalized with respect to the continuum flux can be expressed as  [$T_A - 0.5T_A$(cont)]/0.5$T_A$(cont). The upper panels of Figs.~\ref{H2Oplot} and~\ref{HFplot} show the normalized fluxes versus Doppler velocity in the Local Standard of Rest frame (V$_{\rm LSR}$) for HF (black line) and para-water (green line) toward W49N and W51, respectively. The para-H$_2$O 1$_{11}$--0$_{00}$ ($\nu$=1113.343 GHz) spectra presented here for comparison were observed in the lower sideband of receiver 5a in the same observing campaign as HF. The total on-source integration time was  87.6s for each target. The methods we used to reduce the HF data were also applied to the para-water data (see above).  The dashed lines represent the continuum temperature  $T_A$(cont) (K) normalized to unity and the zero flux level. One can see that toward both sight lines the sideband gain ratios are indeed consistent with unity.

\section{Results}
\begin{figure}
\centering
\includegraphics[width=9cm]{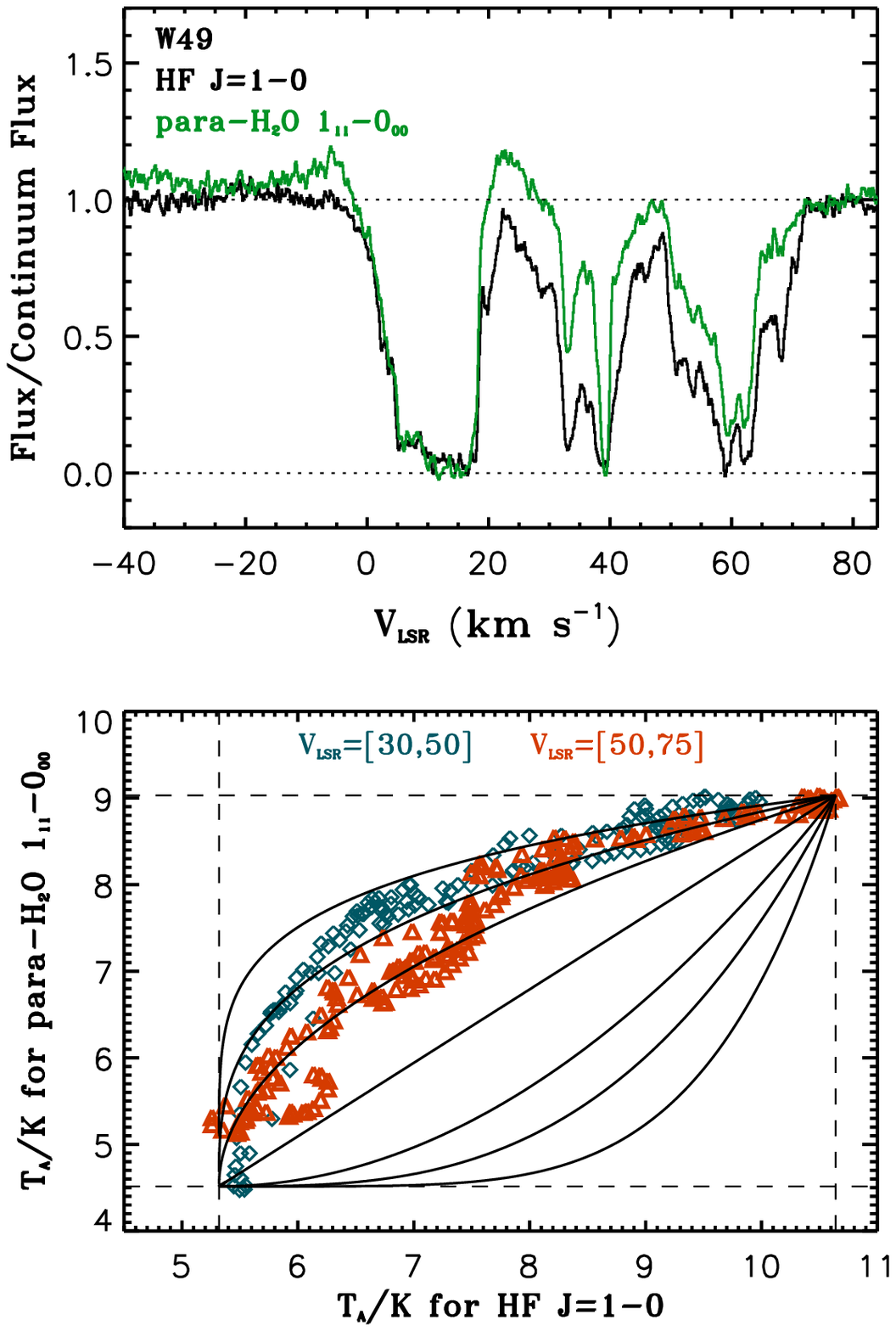}
\caption{Top: Normalized spectra of HF and para-H$_2$O over V$_{\rm LSR}=$ [$-$40,85] km s$^{-1}$. Bottom: HF $J=1-0$ antenna temperature versus para-H$_2$O 1$_{11}$--0$_{00}$ antenna temperature over the velocity ranges indicated at the top of the figure toward the W49N sight line. The solid black lines represent the expected loci for given optical depth ratios of HF to para-water of 5, 3, 2, 1, 0.5, 0.33, 0.2 from top to bottom.} \label{H2Oplot}\end{figure}

\begin{figure}
\centering
\includegraphics[width=9cm]{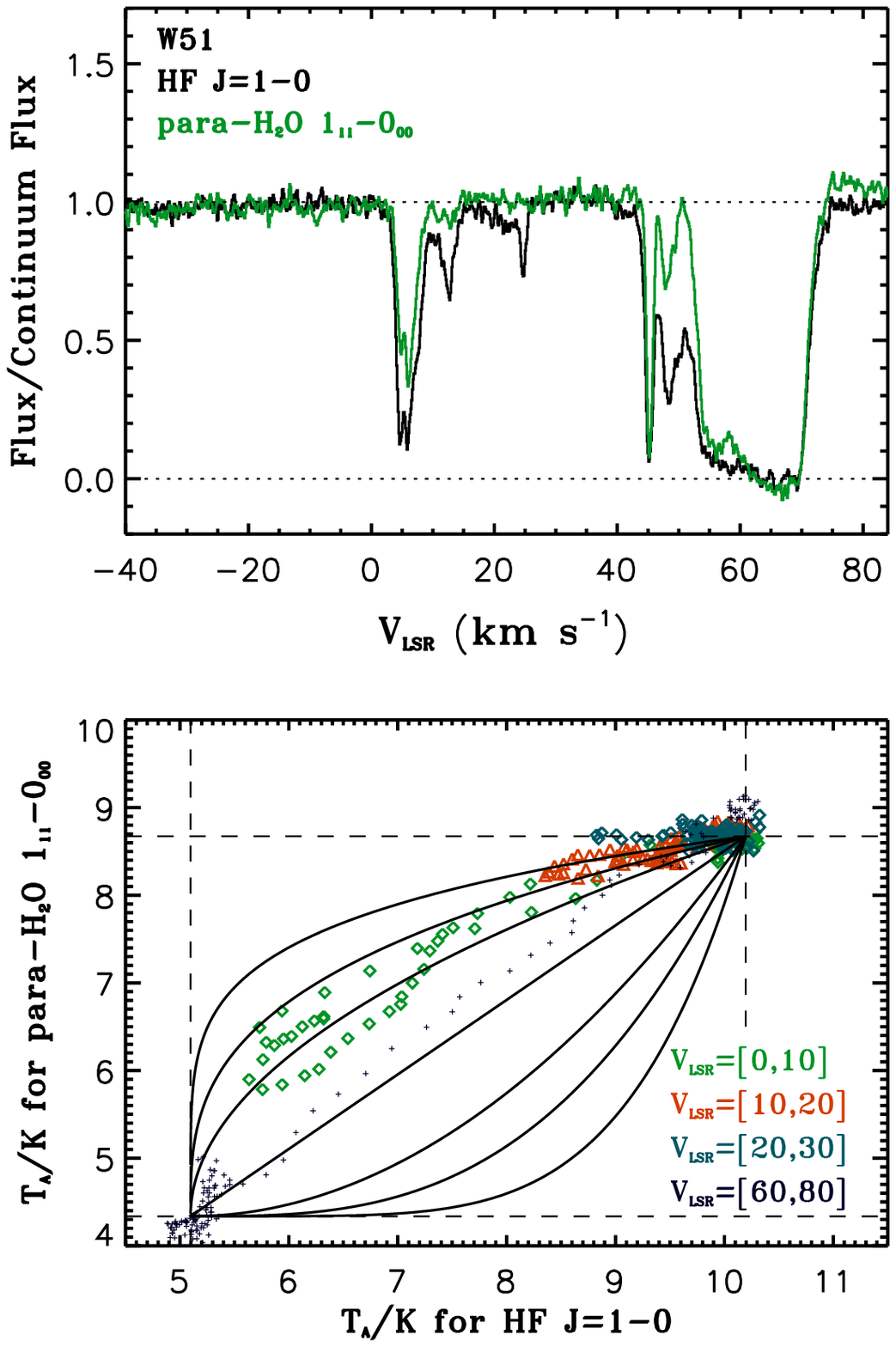}
\caption{Top: Normalized spectra of HF and para-H$_2$O over V$_{\rm LSR}=$ [$-$40, 85] km s$^{-1}$ (see Sect.~2 for details). Bottom: HF $J=1-0$ antenna temperature versus para-H$_2$O 1$_{11}$--0$_{00}$ antenna temperature over the velocity ranges indicated in the lower right corner toward the W51 sight line. Note the absence of para-water absorption compared to HF for V$_{\rm LSR}=$ [20,30] km s$^{-1}$ (blue diamonds). The solid black lines represent the expected loci for given optical depth ratios of HF to para-water of 5, 3, 2, 1, 0.5, 0.33, 0.2 from top to bottom.}\label{HFplot}
\end{figure}

The spectra shown in 
Figs.~\ref{H2Oplot}  and~\ref{HFplot} reveal the presence of HF (black line) and para-water (green line) absorption in a number of foreground clouds along the sight lines to both W49N and W51.
The presence of molecular absorption in these spectra was anticipated by previous observations (e.g., Neufeld et al. 2002; Plume et al. 2004; Godard et al. 2010). 

For W49N, the foreground absorptions in the velocity range $-$10 to $\sim$ 25 km s$^{-1}$ are affected by emission due to the source itself and will not be further discussed here. The two additional sets of HF absorption components in the ranges V$_{\rm{LSR}}=$ 30--50 and 50--75 km s$^{-1}$ are unrelated to the source and have been detected  through absorption from other species such as HCO$^+$, HCN, HNC (Godard et al. 2010), or atomic hydrogen (Fish et al. 2003) and atomic oxygen (Vastel et al. 2000). As for G10.6 -- 0.4 (W31C) observed previously by Neufeld et al.\ (2010), the profiles of HF and para-water are remarkably similar  with the exception of two components at LSR velocities of $\sim$ 68 and 71 km s$^{-1}$ that are clearly much stronger in HF than in water. 

The distributions of foreground material toward W51 also exhibit a one-to-one correspondence between the HF and para-water absorptions with the exception of one component at LSR velocity of 24 km s$^{-1}$ that is clearly detected in HF, but is absent in the para-water spectrum. This component at 24 km s$^{-1}$ has been detected in the absorption spectra of atomic hydrogen (Koo 1997) and the HIFI spectra of CH$^+$ (Falgarone et al. 2010), but -- to our knowledge -- has not been observed in studies of molecular absorption prior to the launch of {\it Herschel}.

The lower panels of Figs.~\ref{H2Oplot} and~\ref{HFplot} compare the HF antenna temperature $T_A$ (K) with the para-water antenna temperature.  Here, each point represents one velocity bin in those LSR velocity ranges exhibiting only moderate HF and para-water optical depths. The dashed lines represent the continuum temperature  $T_A$(cont) (K) and 0.5$T_A$(cont) for HF (vertical dashed lines) and para-water (horizontal dashed lines). The solid black lines in the lower panels of Figs.~\ref{H2Oplot}  and~\ref{HFplot} represent the expected location of the HF and para-water antenna temperatures in each velocity bin for a given optical depth ratio of HF over para-water. From top to bottom, we display HF to para-water ratios of 5, 3, 2, 1, 0.5, 0.33 and 0.2. For G10.6--0.4 (Neufeld et al. 2010),  we find that the optical depth of HF is between 2 and 5 times larger than that of para-water toward both sight lines. Note again the presence of the 24 km s$^{-1}$ component detected in HF and clearly absent in para-water (blue diamonds). 

\section{Discussion}

Toward W51, all para-water components in the LSR velocity range 0 to 30 km s$^{-1}$ are optically thin, while those of HF are either optically thin (the 12 and 24 km s$^{-1}$ components) or moderately thick (the 6 and the 45 km s$^{-1}$   complexes), hence allowing us to directly {\it measure} the column density of these species for each set of clouds.  This stands in contrast to G10.6--0.4, where the large HF $J=1-0$ optical depth allowed Neufeld et al.\ (2010) to obtain only a lower limit on the HF abundance. Note that the HF and para-water lines are optically thick in the velocity range {\bf 50} to 75 km s$^{-1}$, thus only leading to lower limits on their column densities. 

We used a set of multiple Gaussian components to simultaneously fit the HF and para-water profiles in the 0 -- 30 km s$^{-1}$ range. Due to the similarity between the HF and HCO$^+$ profiles in this velocity range, we used the HCO$^+$ sight line cloud decomposition of Godard et al. (2010) as an initial guess in our fits. Under the assumption that both HF and para-water co-exist along the W51 sight line, we fixed the cloud velocity and FWHM of the para-water components to equal  those of HF. 

\begin{figure}
\centering
\includegraphics[width=9cm]{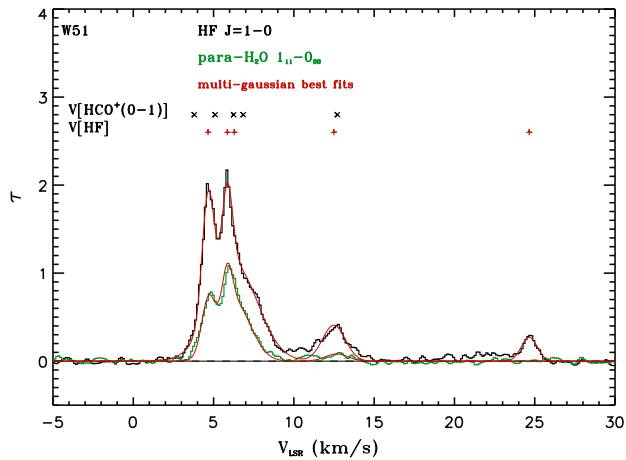}
\caption{Optical depth spectra of HF $J=1-0$  (black line) and para-water 1$_{11}$--0$_{00}$  (green line) toward W51. The ($\rm x$) mark the positions of the HCO$^+$ components (see Godard et al. 2010). The (+) mark the positions of HF components resulting from our multi-component Gaussian fit to the HF data. When fitting the para-water spectrum, the component positions and FWHM were held fixed at the values derived from the HF analysis.} \label{HFfit}
\end{figure}

Figure~\ref{HFfit} displays the optical depth spectra of HF (black line) and para-water (green line) in the $-5$ to 30 km s$^{-1}$ velocity range toward W51, with the best fit to each spectrum superimposed as a red line. The quality of both fits clearly demonstrates that our assumption regarding the similarity in distribution for HF and para-water is justified for the cloud complex at 6 km s$^{-1}$ and the 12 km s$^{-1}$ component. This figure further shows that the distribution of HCO$^+$ does differ slightly from that of HF for the 6 km s$^{-1}$ cloud complex and that the 24 km s$^{-1}$ component is traced by HF alone. The  HCO$^+$ counterpart  of the 45 km s$^{-1}$ complex is blended with strong emission, preventing Godard et al. (2010) from deriving the cloud structure for this absorption complex. In addition the blending between the various features seen in our data is too severe to lead to a unique fitting solution. We therefore integrated the HF and water absorption profiles over the entire 42--47 km s$^{-1}$ range (see Gerin et al. 2010).

Toward W49N, the HF and para-water cloud complexes in the 30--50 and 50--65 km s$^{-1}$ ranges are all optically thicker than those toward W51. Additionally, the various cloud components detected in HCO$^+$ by Godard et al. (2010) are more severely blended than toward W51, preventing us from obtaining a unique fit to the cloud distributions for W49N. Consequently, we integrated the HF and para-water optical depths over the entire velocity ranges with the exception of the two absorbing clouds at LSR velocities of 68 and 71 km s$^{-1}$. These two components are optically thin in both HF and para-water and were fitted with Gaussians using the same method as for W51. 

Following Neufeld et al. (2010), we derived the HF and para-water column densities for each LSR velocity range by dividing the velocity-integrated optical depths of HF and para-water by 4.16\,10$^{-13}\,\rm{cm}^{2}$ / km s$^{-1}$ and 4.30 \,10$^{-13}\,\rm{cm}^{2}$ / km s$^{-1}$, respectively, to obtain the results given in Table~\ref{tbl-1}.  Here, we assume that the absorbing material completely covers the continuum emission region, and that each molecule is primarily in its ground state.  The latter assumption is justified because the gas density is much lower than the critical density at which the collisional deexcitation and spontaneous decay rates would be equal (Neufeld et al.\ 2010). For the same velocity ranges, we have presented estimates for the H$_2$ and atomic H column densities.  In diffuse molecular clouds (also known as ``translucent clouds" in the classification proposed by Snow \& McCall 2006), direct measurements of the H$_2$ and CH column density via far- and near-UV absorption spectroscopy showed that molecular hydrogen and CH trace each other linearly, with $N$(CH)$=$3.5\,10$^{-8}$ $N$(H$_2$) (Sheffer et al. 2008 and references therein). Hence, CH is often used as proxy for H$_2$ when the latter cannot be measured directly. Gerin et al. (2010) detected absorption of CH ($\nu$$=$ 536.761 GHz) toward both W49N and W51. The distribution of the CH absorbing clouds matches that of HF toward both sight lines. We therefore used the CH column densities derived by Gerin et al. (2010)  and the Sheffer et al. (2008) relationship to infer the H$_2$ column densities toward W49N and W51. The uncertainty in our determination of the H$_2$ column density is dominated by the scatter in the CH-H$_2$ relationship, estimated by Sheffer et al. as 0.21 dex, corresponding to a factor of 1.6. The atomic hydrogen column densities for W51 were obtained from the 21 cm absorption spectra obtained by Koo (1997) for an assumed 21 cm spin temperature of 100~K, while those for W49N were obtained by Godard et al. (2010), based upon 21 cm observations presented by Fish et al. (2003).

In their analysis of the chemistry of fluorine-containing molecules in the interstellar medium, Neufeld \& Wolfire (2009) predicted  (e.g. their Fig.~7) that over a wide range of conditions the HF and H$_2$ column densities would track each other exactly, with the ratio $N({\rm HF})/2N({\rm H_2})$ equal to the gas-phase elemental abundance of F relative to H nuclei.
Given the average gas-phase abundance of interstellar fluorine in diffuse atomic gas of $N_{\rm F}$/$N_{\rm H}$$=$ 1.8 \,10$^{-8}$ (Snow, Destree \& Jensen 2007), this result implies a $N({\rm HF})/N({\rm H}_2)$ ratio of 3.6 \,10$^{-8}$.  Given  the uncertainty in the $N({\rm CH})-N({\rm H}_2)$ relationship used to infer the molecular hydrogen column densities and the variations in the gas-phase fluorine abundance, the results shown in  Table~\ref{tbl-1} are consistent with that prediction.  The lack of correlation between $N({\rm HF})$ and $N({\rm H})$ for the cloud components presented in this work also suggests that HF is absent in purely atomic material, but has the potential to serve as an excellent tracer of H$_2$.  Hydrogen fluoride provides a sensitive probe of clouds of small H$_2$ column density.  Indeed, the observations of HF reported here reveal a low column density molecular cloud along the W51 sight line, at an LSR velocity of $\sim 24\, \rm km\,s^{-1}$, which had not been identified in molecular absorption line studies prior to the launch of {\it Herschel}. 
  
\begin{table}[ht]
\caption{Summary of derived column densities and abundances.\label{tbl-1}}
\begin{tabular}{cccccc}
\hline

\\
V$_{\rm LSR}$ & $N$(HF) & $N$(H$_2$O)  & $N$(H) $^a$ & $N$(H$_2$) $^b$ &HF/H$_2$ \\
(km/s) & $\rm 10^{12}\,cm^{-2}$ & $\rm 10^{12}\,cm^{-2}$ & $\rm 10^{20}\,cm^{-2}$ & $\rm 10^{20}\,cm^{-2}$ & \\
\\
 \hline
\\
\multicolumn{5}{l}{Results for W51} \\
\\
0--10  & 14.5$\pm$1.1 & 6.2$\pm$0.5 &  13.9 & 10.5& $1.4\,10^{-8}$  \\
10--20 & 1.8$\pm$0.3 & 0.4$\pm$0.1 & 6.4& 1.4 & 1.3 $\,10^{-8}$\\
20--30 & 0.8$\pm$0.1 & $<$0.2 & 11.1 & $<$0.7 & $>$1.2$\,10^{-8}$\\ 
42--47  & 8.0$\pm$1.0 & 5.4$\pm$0.7 & ... & 5.0 &   1.6$\,10^{-8}$ \\
50--75 &  $>$130 & $>$113 & 22.1 &  ... &  ... \\
\\
\multicolumn{5}{l}{Results for W49N} \\
\\
30--50 & 55$\pm$10 & 22$\pm$8 & 69.5 & 37 & 1.5$\,10^{-8}$\\
50--78 & 69$\pm$10 & 34$\pm$9 & 72.3 & 66& 1.1$\,10^{-8}$\\
67--71 & 5.6$\pm$1.0 &  1.5$\pm$0.3& ... & 4.4& 1.3$\,10^{-8}$ \\
\\
\hline 
\\
\end{tabular}
\tablefoottext{a}{from Koo 1997 (W51) and Godard et al.\ 2010 (W49N), for an assumed 21 cm spin temperature of 100~K.}\\
\tablefoottext{b}{Derived from CH observations (Gerin et al.\ 2010), assuming $N$(CH)= 3.5$\times$10$^{-8}$ $N$(H$_2$) (Sheffer et al. 2008). The relationship shows a scatter of 0.21 dex, corresponding to a factor of 1.6.}\\

\end{table}

\begin{acknowledgements}

HIFI has been designed and built by a consortium of institutes and university departments from across
Europe, Canada and the United States under the leadership of SRON Netherlands Institute for Space
Research, Groningen, The Netherlands and with major contributions from Germany, France and the US.
Consortium members are: Canada: CSA, U.~Waterloo; France: CESR, LAB, LERMA, IRAM; Germany:
KOSMA, MPIfR, MPS; Ireland, NUI Maynooth; Italy: ASI, IFSI-INAF, Osservatorio Astrofisico di Arcetri-
INAF; Netherlands: SRON, TUD; Poland: CAMK, CBK; Spain: Observatorio Astron\'omico Nacional (IGN),
Centro de Astrobiolog\'a (CSIC-INTA). Sweden: Chalmers University of Technology - MC2, RSS \& GARD;
Onsala Space Observatory; Swedish National Space Board, Stockholm University - Stockholm Observatory;
Switzerland: ETH Zurich, FHNW; USA: Caltech, JPL, NHSC. M.S. acknowledges support from grant N 203 393334 from Polish MNiSW.

This research was performed in part through a JPL contract funded by the National Aeronautics and Space Administration.
     
\end{acknowledgements}


\begin{thebibliography}{}

\bibitem[]{de Graauw} de Graauw, T., Helmich, F., Phillips, T.  et al. 2010, A\&A, in press ({\it Herschel} special issue)

\bibitem[2010]{edith} Falgarone, E., Godard, B., Cernicharo, P. et al., this volume

\bibitem[2003]{fish} Fish, V. L., Reid, M. J., Wilner, D. J., \& Churchwell, E. 2003,
      ApJ, 587, 701
      
\bibitem[2010]{godard} Godard, B., Falgarone, E., Gerin, M., Hily-Brant, P., \& De Luca, M. 2010,
      A\&A, submitted

\bibitem[]{gerin} Gerin, M.  et al., this volume


\bibitem[1992]{gwinn} Gwinn, C. R., Moran, J. M, \& Reid, M. J. 1992,
ApJ, 393, 149

\bibitem[]{kirk} Kirk et al., 2010, A\&A, in press ({\it Herschel} special issue)

\bibitem[1997]{koo} Koo Bon-Chul 1997, ApJS, 108, 489

 \bibitem[2002]{neufeld} Neufeld, D. A., Kaufman, M. J., Goldsmith, P. F., Hollenbach, D. J, \& Plume, R.  2002,
      ApJ, 580, 278
      
 \bibitem[2005]{neufeld} Neufeld, D. A., Wolfire, M.,\& Schilke, P.  2005,
      ApJ, 628, 260

 \bibitem[2009]{neufeld} Neufeld, D. A., \& Wolfire, M. 2009, 
      ApJ, 706, 1594
      
 \bibitem[]{neufeld} Neufeld, D. A. et al., 2010, A\&A, in press ({\it Herschel} special issue) 
      
 \bibitem[1987]{nolt} Nolt, I. J. et al.1987,
   Journal of Molecular Spectroscopy, 125, 274
   
  \bibitem[2010]{Ott} Ott, S. 2010, in ASP Conference Series, Astronomical Data Analysis Software and Systems XIX, Y. Mizumoto, K.-I. Morita and Ohishi, eds, in press 

 \bibitem[]{Phillips} Phillips, T.G. et al., 2010, A\&A, in press  

 \bibitem[]{Pilbratt} Pilbratt, G. et al., 2010, A\&A, in press  
 
\bibitem[]{plume} Plume, R. et al. 2004,
ApJ, 605, 247
      
\bibitem[2010]{sato} Sato, M., Reid, M. J., Brunthaler, A. ,\& Menten, K.M.  2010
      ApJ, submitted 

   \bibitem[2008]{sheffer} Sheffer, Y. et al.,  2008,
      ApJ, 687, 1075 
     
 \bibitem[2007]{snow} Snow, T. P., Destree, J. D., \& Jensen, A. G. 2007,
      ApJ, 655, 285
      
 \bibitem[2006]{snow} Snow, T. P.,  \& McCall, B. J. 2006,
      ARA\&A, 44, 367
\bibitem[2000]{vastel} Vastel, C., Caux, E., Ceccarelli, C. et al. 2000, A\&A 357, 994
\end{thebibliography}
\end{document}